# Prediction of long and short time rheological behavior in soft glassy materials

A. Shahin and Yogesh M Joshi*

Department of Chemical Engineering, Indian Institute of Technology Kanpur,

Kanpur 208016. INDIA.

* E-Mail: joshi@iitk.ac.in

## ABSTRACT

We present an effective time approach to predict long and short time rheological behavior of soft glassy materials from experiments carried out over practical time scales. Effective time approach takes advantage of relaxation time dependence on aging time that allows time-aging time superposition even when aging occurs over the experimental timescales. Interestingly experiments on variety of soft materials demonstrate that the effective time approach successfully predicts superposition for diverse aging regimes ranging from sub-aging to hyper-aging behaviors. This approach can also be used to predict behavior of any response function in molecular as well as spin glasses.

Glasses are out of equilibrium materials [1]. Because of structural arrest these materials cannot explore the phase space over practical time scales [2]. However, their structure evolves and progressively lowers its energy in search of a global minima [3]. Struik [4] termed this phenomenon as physical aging which accompanies slowing down of the relaxation dynamics with time elapsed since physical arrest. While analyzing the physical behavior and/or designing the glassy materials for strategic applications the biggest hurdle such an out of equilibrium dynamics poses is deviation from time - translational invariance (TTI) [5]. This problem imparts limited applicability of time temperature superposition and Boltzmann superposition principle in its most conventional form [4, 5] limiting the predictive capacity in time and frequency domain. In this letter we propose an effective time procedure for an important class of nonergodic materials: soft glassy materials, to predict a long and a short time rheological behavior from experiments performed over practical time scales.

Over a past decade, soft materials such as pastes, creams, concentrated emulsions, suspensions, gels, foams, paints, slurries and variety of food materials that show out of equilibrium dynamics have acquired significant academic and commercial importance. Their time dependent rheological behavior and its complex dependence on deformation field have posed strong challenges in their industrial processing. In addition, these materials, also known as soft glassy materials, show aging behavior similar to that of molecular glasses and spin glasses wherein their relaxation time increases as a function of time [2, 6-8]. In a seminal contribution, Struik [4] proposed a procedure to predict long time creep behavior of polymeric glasses from a superposition of creep curves obtained through momentary experiments. Similar to Struik's attempt for polymeric glasses, time aging-time superposition has been applied to soft glassy materials and spin glasses to obtain dependence of relaxation time on age by many groups [6, 9, 10]. However most of these efforts investigate only momentary rheological behavior (process time << age), suggesting a need for a procedure to predict a behavior over long process times.

In this work we have used an aqueous suspension of 2.8 wt. % Laponite RD (Southern Clay Products, Inc.) with 0.5 wt. % Polyethylene oxide (PEO, molecular weight=6000, Loba Chemie). We have also used two commercial soft glassy materials: hair gel [11] and water based emulsion paint [12]. Hair gel is a copolymer based gel containing hydrophilic and hydrophobic ingredients while an emulsion paint is a concentrated emulsion of acrylic based polymer in an aqueous continuous phase. Procedure to prepare aqueous Laponite-PEO suspension is discussed elsewhere [10, 13]. We used a stress controlled rheometer AR 1000 for rheological studies. Experiments on Laponite suspension and emulsion paint were carried in a couette geometry (bob diameter 28 mm with gap 1 mm) while that of on hair gel was carried out in a cone and plate geometry (diameter 40 mm with cone angle 2°). In a typical experimental protocol, Laponite suspension on a certain date since preparation was loaded in a couette cell and shear melted for 20 minutes by applying oscillatory shear stress with magnitude 80 Pa and frequency 0.1 Hz. After shear melting, sample was allowed to age for a predetermined waiting time: $t_w$ and subsequently creep experiments were carried out at 5 Pa. During an aging period a small amplitude oscillatory shear with stress magnitude of 1 Pa and frequency of 0.1 Hz was applied. For a hair gel we carried out shear melting at 300 Pa oscillatory stress and creep experiments at 20 Pa. For an emulsion paint shear melting was carried out at 200 Pa oscillatory stress and creep experiments were carried out at 3 Pa. During a course of experiments, a free surface was covered with a thin layer of low viscosity silicon oil to avoid drying and/or evaporation of water. All the experiments were carried out at 25°C.

All the three samples studied in this work show time dependent evolution of its viscoelastic properties. Under such circumstances system does not obey TTI, such that linear viscoelastic response is not just a function of the difference between the total time $t$ and the time at which deformation was applied $t_w$: $(t - t_w)$ as is the case when TTI is applicable; but is also independently depends on $t_w$ [5, 14]. The rheological response in such situations can be captured by replacing the time difference $t - t_w$ by difference in the effective times $\xi(t) - \xi(t_w)$, where the effective time scale $\xi(t)$ is given by [1, 5, 15]:

$$\xi(t) = \tau_0 \int_0^t \frac{dt'}{\tau(t')}, \tag{1}$$

where $\tau$ is a relaxation time of a system while $\tau_0$ is a microscopic relaxation time. The difference in effective time is:

$$\xi(t) - \xi(t_w) = \tau_0 \int_{t_w}^{t} \frac{dt'}{\tau(t')}. \qquad (2)$$

In a stress controlled deformation field such as creep, strain induced in the material is given by:

$$\gamma_{12}(t) = \int_{-\infty}^{t} J\left(\xi(t) - \xi(t_w)\right) \frac{d\sigma_{12}}{dt_w} dt_w = \int_{-\infty}^{t} J\left(\tau_0 \int_{t_w}^{t} \frac{dt'}{\tau(t')}\right) \frac{d\sigma_{12}}{dt_w} dt_w. \qquad (3)$$

In a limit $t - t_w << t_w$, where evolution of relaxation time over duration of creep time $t - t_w$ can be neglected: $\tau(t) \approx \tau(t_w)$, we get $\xi(t) - \xi(t_w) = \tau_0(t - t_w)/\tau(t_w)$. Struik [4] was the first to propose a time – aging time superposition procedure wherein he considered a rheological response at different aging times $t_w$ only in a limit of $t - t_w << t_w$. The corresponding shifting on time axis to get a superposition yielded a dependence of relaxation time on aging time: $\tau(t_w)$. To guarantee $\tau(t) \approx \tau(t_w)$, Struik [4] suggested a protocol that the process time $t - t_w$ considered in superposition should be at most 10 % of age ($t_w$). However if dependence of relaxation time on aging time is stronger than linear, consideration of 10 % of aging time does not guarantee absence of aging during an experiment. Under such circumstances, it is prudent to plot rheological response function as a function of $\int_{t_w}^{t} \frac{dt'}{\tau(t')}$ to get a superposition over an entire domain of process times. However in such case, one needs to assume a functional form of the dependence: $\tau(t')$ a priory. Although, we have considered only a single mode of relaxation in eq. 3, any real material possesses distribution of modes. Therefore, time - aging time superposition is possible only if all the modes age similarly preserving the shape of a spectrum of relaxation times [4, 8].

In order to facilitate time – aging time superposition we carried out the creep experiments on the Laponite suspension on 13th day and 30th day after preparation. Laponite suspension is known to undergo irreversible aging over very long durations [13]. Therefore Laponite suspension on 13th day and on 30th day should be considered as two independent materials. In an upper inset of figure 1 we have plotted creep compliance as a function of creep time $(t - t_w)$, for the experiments carried out on day 13 at various $t_w$ in a range 1800 s to 14400 s. On day 13 the shear melted suspension was in a liquid state $(G'' > G')$ over a duration of the aging times explored in this work. It can be seen that lesser strain gets induced in the samples for the experiments carried out at greater ages. A superposition is expected when vertically shifted transient compliance $J(t)G(t_w)$ is plotted as a function of $\int_{t_w}^{t} \frac{dt'}{\tau(t')}$. Vertical shift factor $G(t_w)$ is an elastic modulus at age $t_w$ and superposition fixes its value based upon small time strain induced in the material. For a creep data shown in figure 1, exponential dependence: $\tau(t') = \tau_0 \exp(\alpha t')$ is needed for obtaining a superposition, which leads to difference between effective times given by:

$$\xi(t) - \xi(t_w) = \int_{t_w}^{t} \exp(-\alpha t') dt' = \left[ \exp(-\alpha t_w) - \exp(-\alpha t) \right] / \alpha . \qquad (4)$$

It can be seen that all the creep curves superpose for an entire duration of creep experiments for the assumed exponential dependence. Usually a dependence of relaxation time on aging time faster than linear is termed as hyper aging. Exponential dependence therefore represents a hyper aging dynamics and is successfully modeled by the effective time theory.

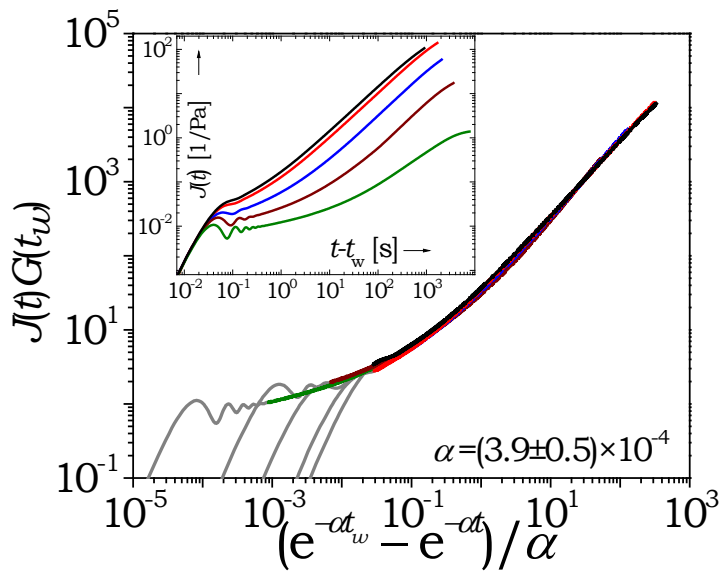

**Figure 1.** Time - aging time superposition for creep curves generated on day 13. An inset shows creep curves obtained at various waiting times (from top to bottom: 1800 s, 3600 s, 7200 s, 10800s 14400 s).

Interestingly for experiments carried out on 30th day after preparation, Laponite suspension entered a glasslike state $(G' > G'')$ immediately after shear melting. In this state relaxation time of most glassy systems such as colloidal, molecular and spin glasses demonstrate a power law dependence on aging time $\left( \tau(t') = \tau_0^{1-\mu} t'^{\mu} \right)$. The power law coefficient $\mu$, which represents $d\ln\tau/d\ln t_w$ has recently been associated with reference average energy well depth at the beginning of aging normalized by $kT$ [10]. With power law dependence, eq. 2 leads to [5]:

$$\xi(t) - \xi(t_w) = \tau_0 \int_{t_w}^{t} \frac{dt'}{\tau_0^{1-\mu} t'^{\mu}} = \tau_0^{\mu} \left\{ \frac{\left( t^{1-\mu} - t_w^{1-\mu} \right)}{1-\mu} \right\} = \tau_0^{\mu} \theta(t, t_w). \qquad (5)$$

In an inset of figure 2, we have plotted creep curves obtained on day 30, which have qualitatively similar features as shown in an inset of figure 1. Figure 2 shows that the creep curves demonstrate a remarkable superposition for an entire duration of creep experiments, when transient compliance normalized by modulus $J(t)G(t_w)$ is plotted as a function of $\theta(t, t_w)$, for $\mu = 5.16$. Value of $\mu$ greater than unity also describes a hyper aging regime.

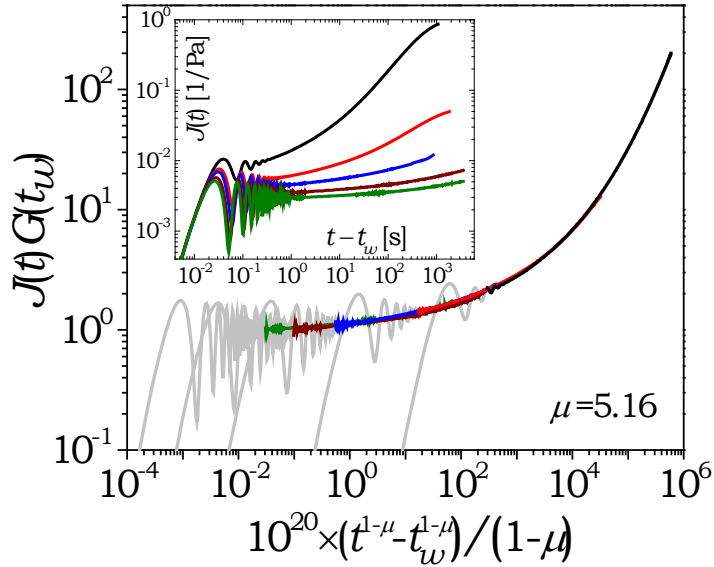

**Figure 2.** Time - aging time superposition for creep curves generated on day 30. An inset shows creep curves obtained at various waiting times (from top to bottom: 1800 s, 3600 s, 7200 s, 10800s 14400 s).

In order to test a validity of effective time procedure we studied two very diverse commercial soft glassy materials hair gel and emulsion paint. Figure 3 shows that when vertically shifted compliance is plotted against $\theta(t, t_w)$ (eq. 5) creep curves demonstrate an excellent superposition irrespective of a duration of a creep time. Hair-gel shows hyper-aging behavior ($\mu > 1$) while paint shows sub-aging behavior ($\mu < 1$). This observation shows applicability of this approach to very diverse types of soft materials.

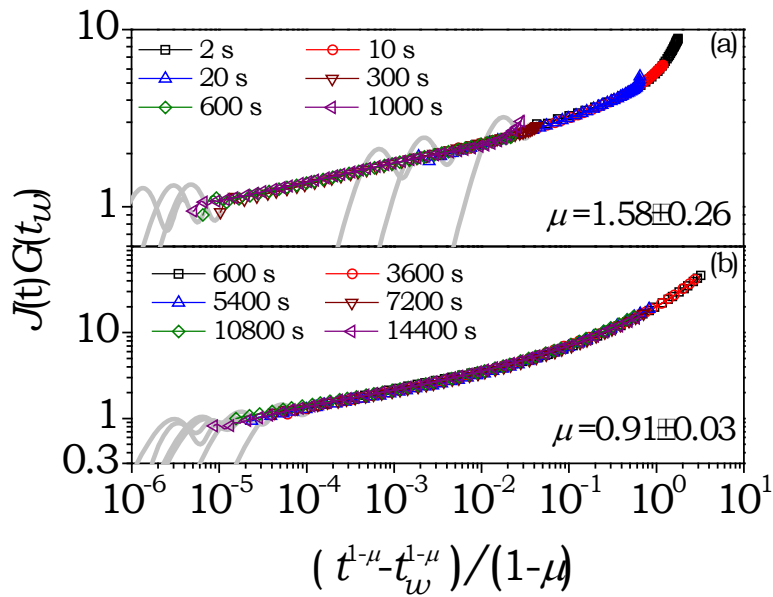

**Figure 3.** Time-aging time superposition for two commercial soft materials (a) hair gel and (b) acrylic emulsion paint.

The validation of effective time procedure to obtain superposition also aids to predict long term or very short term creep behavior of glassy materials from tests carried out over accessible timescales. The effective time associated with the superposition shown in figure 2 can be inverted to real time to obtain creep curves at individual aging times. For example, for any point in superposition, horizontal axis coordinate $\theta$ can be inverted to get real creep time $t - t_w$ as,

$$t - t_w = \left\{ \theta(1-\mu) + t_w^{1-\mu} \right\}^{1/(1-\mu)} - t_w. \tag{6}$$

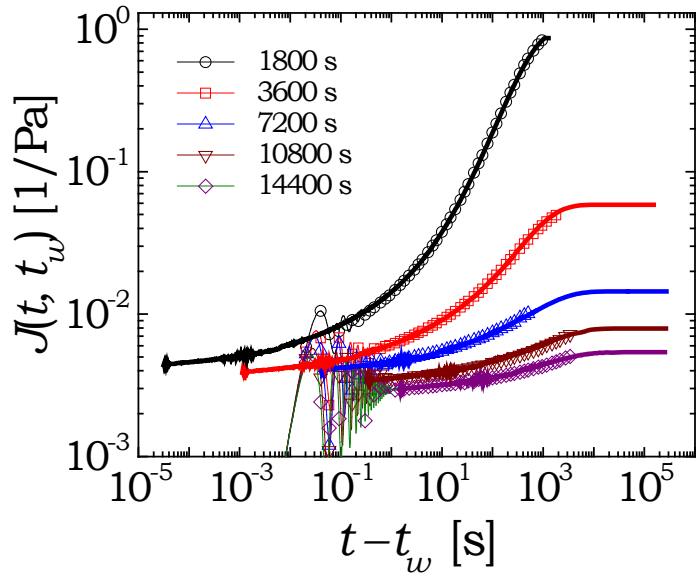

**Figure 4.** Prediction of creep behavior from superposition shown in figure 2 and equation 6. The points represent experimental data shown in an inset of figure 2 while the thick lines represent predicted creep curves.

In figure 4 we have plotted experimental creep curves along with the prediction of equation 6 for various ages $t_w$ and for $\mu = 5.16$. It can be seen that equation 6 on one hand predicts a long time behavior of the samples having large ages, while on the other hand predicts very short time behavior of the samples having small ages. This procedure, therefore remarkably allows prediction of long time behavior by carrying out few short time duration experiments. Equivalently the prediction of very short time behavior also has great significance because, such small times are never accessible in the experiments due to limitations of instrument inertia and/or data acquisition speeds. Eqs. 5 and 6 suggest that the predictive capacity of this procedure decreases with decrease in $\mu$ and in the limit of $\mu \to 0$, real time and effective time become equal to each other. While carrying out the prediction of the rheological behavior, however, it is necessary to ensure that dependence of relaxation time on aging time remains identical to that considered in analysis of superposition over the time scales of prediction. For $\mu > 1$, term in the braces of Eq. 6 becomes negative for $\theta > t_w^{1-\mu}/(\mu-1)$. This suggests that for $\mu > 1$, compliance in the creep experiments will always attain a plateau at high creep times with its value equal to corresponding value of compliance at $\theta = t_w^{1-\mu}/(\mu-1)$ from the superposition. It can be seen that figure 4 indeed predicts compliance to show a plateau.

There are several important features associated with the observed behavior. This work is the first to predict very long time and very short time rheological behavior in soft glassy materials. This procedure can be directly used to predict long and short time behavior of any other response function associated with glassy materials undergoing aging (such as relaxation modulus and compliance in molecular glasses, thermoremanent magnetization in spin glasses, etc.). This work shows that prediction can be obtained in diverse type of glassy materials in a broad range of aging regimes: from sub-aging to hyper-aging dynamics. Interestingly, this work also demonstrates time – aging time superposition on effective time axis for non power law (exponential) dependence of relaxation time on aging time for system in a liquid state. In addition, superposition carried out using this procedure is not constrained by Struik protocol.

In conclusion, an effective time procedure is presented, which takes into account dependence of relaxation time on the aging time, to facilitate a superposition of creep data obtained at different ages. Using very diverse soft glassy materials we demonstrate time - aging time superposition for wide dependences of relaxation time on aging time such as an exponential and power law $\left( \tau(t_w) = \tau_0^{1-\mu} t_w^{\mu} \right)$ with coefficient $\mu$ varying from sub aging regime ($\mu < 1$) to hyper aging regime ($\mu > 1$). We also demonstrate how the superposition obtained using effective time theory can be used to predict very small time and very large time rheological behavior using experimental data obtained over practically measurable time scales. This procedure can also be employed to predict short and long term behavior of response functions in molecular and spin glasses.

Acknowledgement: We thank Professor Mike Cates for his comments and useful discussions during YMJ's visit to Edinburgh, funded by the Royal

Society of Edinburgh - Indian National Science Academy International Exchange Programme. This work is supported by the department of science and technology, Government of India under IRHPA scheme.

------------------------